\newcommand{\NoOfSources}{24}  
\newcommand{\NoOfKeeps}{19}  
\newcommand{\NoIterations}{$10^5$} 
\newcommand{\RmsStack}{$\sigma_{\tau}=2.6\times 10^{-4}$} 
\newcommand{\TimesBetter}{4.2} 
\newcommand{\EW}{$0.018\,\rm km\,s^{-1}$} 
\newcommand{\MonteCarloEW}{$\left<\rm{EW}\right> = 0.0182^{+0.0044}_{-0.0036}\,\rm km\,s^{-1}$} 
\newcommand{\MCfwhm}{$\rm{FWHM} = 50^{+15}_{-7}\,\rm km\,s^{-1}$} 
\newcommand{\MCpeak}{$\tau_0 = 3.0^{+1.0}_{-0.4}\times10^{-4}$} 
\newcommand{\StackedSNR}{5} 
\newcommand{\StackedSpinTemp}{$\left<T_s\right> = 7200^{+1800}_{-1200}\,\rm K$}
\newcommand{\LowLimit}{$4300\rm\,K$}

\documentclass{aastex}
\usepackage{emulateapj5}

\newif\ifAMStwofonts
\AMStwofontstrue

\def\ga{\mathrel{\hbox{\rlap{\hbox{\lower4pt\hbox{$\sim$}}}\hbox{$>$}}}}
\def\la{\mathrel{\hbox{\rlap{\hbox{\lower4pt\hbox{$\sim$}}}\hbox{$<$}}}}

\begin{document}

\title{Excitation Temperature of the 
       Warm Neutral Medium as a New Probe of 
       the Lyman-$\alpha$ Radiation Field}

\author{Claire E. Murray\altaffilmark{1}, 
        Robert R. Lindner\altaffilmark{1}, 
        Sne\v{z}ana Stanimirovi\'{c}\altaffilmark{1},
        W. M. Goss\altaffilmark{2}, 
        Carl Heiles\altaffilmark{3}, 
        John Dickey\altaffilmark{4}, 
        Nickolas M. Pingel\altaffilmark{1}, 
        Allen Lawrence\altaffilmark{1},
        Jacob Jencson\altaffilmark{2,5}, 
        Brian L. Babler\altaffilmark{1}, and
        Patrick Hennebelle\altaffilmark{6},}

\altaffiltext{1}{Department of Astronomy, University of 
                 Wisconsin, Madison, WI 53706, USA}
\altaffiltext{2}{National Radio Astronomy Observatory,
                 P.O. Box O, 1003 Lopezville, Socorro, 
                 NM 87801, USA}
\altaffiltext{3}{Radio Astronomy Lab, UC Berkeley,
                 601 Campbell Hall, Berkeley, CA 94720, USA}
\altaffiltext{4}{University of Tasmania, School of Maths 
                 and Physics,Private Bag 37, Hobart, 
                 TAS 7001, Australia}
\altaffiltext{5}{Department of Astronomy, 
		Ohio State University, 140 West
		18th Avenue, Columbus, OH 43210}
\altaffiltext{6}{Laboratoire AIM, Paris-Saclay, 
		CEA/IRFU/SAp$-$CNRS$-$ Universit\'e
		Paris Diderot, 91191 Gif-sur-Yvette Cedex, France}

\begin{abstract}
We use the Karl G. Jansky Very Large Array (VLA) to conduct a high-sensitivity 
survey of neutral hydrogen (H\textsc{i}) absorption
in the Milky Way. In combination with corresponding
H\textsc{i} emission spectra obtained mostly with the Arecibo
Observatory,  we detect a widespread warm neutral medium (WNM) component with
excitation temperature \StackedSpinTemp{}
(68\% confidence).
This temperature lies above theoretical predictions 
based on collisional excitation alone, implying that Ly-$\alpha$
scattering, the most probable additional source of excitation, 
is more important in the interstellar medium (ISM) than
previously assumed. Our results demonstrate that
H\textsc{i} absorption can be used to constrain the
Ly-$\alpha$ radiation field, a critical quantity
for studying the energy balance in
the ISM and intergalactic medium yet
notoriously difficult to model because of its
complicated radiative transfer, in and around
galaxies nearby and at high redshift.
\end{abstract}


\section{Introduction}

Understanding physical conditions within the diffuse 
neutral interstellar medium (ISM) is essential for 
producing realistic models of star and galaxy formation. 
Ambient gas temperature and density are crucial input 
parameters for heating, cooling and feedback recipes 
on all astronomical scales.  While numerical simulations 
are becoming increasingly complex, details regarding neutral 
gas temperature distributions, shielding properties, 
essential feedback sources, and excitation processes 
are still very much under debate \citep{Bryan07,Christensen12}.
The excitation processes of the 21-cm line are especially
important for interpreting radio signals from early
epochs of cosmic structure formation (e.g., the cosmic
dark ages and subsequent epoch of reionization), when
neutral hydrogen dominated the baryonic content of the 
Universe and facilitated the formation of the first stars 
and galaxies \citep{Pritchard12}.  To interpret 21-cm 
signals from the early universe, it is necessary to 
decouple astrophysical effects from cosmological effects 
which is likely best done by analyzing excitation processes 
in the local ISM. 

Traditional ISM models contain two neutral phases, the 
cold neutral medium (CNM) and the warm neutral medium (WNM), 
individually in thermal and pressure 
equilibrium \citep{Field69,MO77,Wolfire03}.
Widely-accepted theoretical properties of these phases in 
the Milky Way include: a kinetic temperature of
$T_k\sim40$--$200\rm\,K$ and a volume density of 
$n(\rm H\textsc{i})\sim5$--$120\rm\,cm^{-3}$
for the CNM, and $T_k\sim4100$--$8800\rm\,K$ and 
$n(\rm H\textsc{i})\sim0.03$--$1.3\rm\,cm^{-3}$ for 
the WNM \citep{Wolfire03}. 

A convenient tracer for neutral gas is the 
H\textsc{i} 21-cm line, originating from 
the hyperfine energy splitting caused by magnetic moment 
interactions between the hydrogen atom's electron and proton.
The high optical depth of the CNM makes 21-cm absorption 
signatures easy to detect, even with low sensitivity 
observations \citep[e.g., ][]{Lazareff75,Dickey77,Crovisier78,PDST78,Dickey83,BW92,HT031,K03,Mohan04,Roy06,Begum}. 
The excitation temperature (or spin temperature, $T_s$)
of the CNM can be directly estimated by solving radiative 
transfer equations.

In contrast, the WNM is characterized by very-low 
peak optical depth and so measuring its 
spin temperature from absorption requires extremely high 
sensitivity, $\sigma_{\tau}\leq 10^{-3}$ and attention to systematic errors. 
Only two direct measurements of WNM spin temperature exist 
so far \citep{CDG98,DCG02}. However, \citet{CDG98} observed 
absorption in the direction of Cygnus A, an exceptionally 
bright radio continuum source with a flux density of 
$\sim 400\rm\,Jy$, and achieved excellent sensitivity with 
orders of magnitude less integration time than is required for 
the majority of ($>3\rm\,Jy$) radio continuum sources.  Less 
accurate are upper limits on $T_k$ (and thus $T_s$ since $T_s\leq T_k$) estimated by spectral 
linewidths \citep{Mebold82,K03,HT031,Roy13two}, and approximate
$T_s$ estimates obtained by assigning a single temperature 
to strongly-absorbing, complex H\textsc{i} profiles 
\citep{KBR11,Roy13one}.

Relating measured H\textsc{i} spin temperatures to 
model-predicted kinetic temperatures is further complicated by 
the uncertainty in the excitation mechanisms involved.
The 21-cm transition in the high-density CNM is expected to be 
thermalized by collisions with electrons, ions, and other 
H\textsc{i} atoms, resulting in $T_s \sim T_k$.  In contrast, 
low densities in the WNM imply that collisions cannot thermalize 
the 21-cm transition and therefore $T_s<T_k$ 
\citep{Field58,DW85,Liszt01}. However, the Ly-$\alpha$ radiation 
field from Galactic and extragalactic sources can serve to 
thermalize the transition.  This requires a very large optical 
depth and a large number of scatterings of Ly-$\alpha$ photons
to bring the radiation field and the gas into local thermal 
equilibrium.  While the underlying atomic physics is 
understood \citep{Wouthuysen52,Field58,Pritchard12}, the details of Ly-$\alpha$ 
radiative transfer are complicated and depend on the topology 
and the strength of the Ly-$\alpha$ radiation field, which are 
complex and poorly constrained in the multi-phase 
ISM \citep{Liszt01}. 

To measure the physical properties of the WNM and the 
coupling between neutral gas and Ly-$\alpha$ radiation, 
we are conducting a large survey, 
21-cm Spectral Line Observations of Neutral Gas with the 
EVLA (21\,SPONGE), to obtain high-sensitivity Milky 
Way H\textsc{i} absorption spectra using the Karl G. 
Jansky Very Large Array (VLA).  The recently upgraded 
capabilities of the VLA allow us to routinely achieve 
RMS noise levels in optical depth of $\sigma_{\tau}\sim7\times10^{-4}$ 
per 0.42 km\,s$^{-1}$ channel, which are among the most 
sensitive observations of H\textsc{i} absorption to date. 
Currently, \NoOfSources{} out of 58 sightlines are complete 
after over 200 hours of observing time.
This paper summarizes our initial results from this project 
and the detection of the WNM using a newly developed analysis 
technique based on the spectral stacking of H\textsc{i} absorption and 
emission spectra. In Section~\ref{s:obs}, we summarize our 
observing and data processing strategies, results of the 
stacking analysis are provided in  Section~\ref{s:stacking} 
and discussed in Section~\ref{s:discussion}, and we present
our conclusions in Section~\ref{s:conclusions}.

\section{Observations and Data Processing}
\label{s:obs}

\subsection{Observations}

Each source, selected from the NRAO/VLA Sky Survey \citep{NVSS}, 
was chosen to have a 1.4 GHz flux density $\geq 3\,\rm Jy$ to avoid
excessively long integration times with the VLA to reach the 
desired sensitivity.  In addition, we select sources generally 
at high Galactic latitude ($|b|$ $>$ 10$^\circ$) to avoid 
complicated CNM profiles associated with the Galactic plane,
and with angular sizes less than $1^\prime$ to avoid resolving 
substantial flux density.  All VLA observations use three separate, 
standard L-band configurations, each with one dual-polarization 
intermediate frequency band of width $500\,\rm kHz$ with 
256 channels, allowing for a velocity coverage of 
$107.5\,\rm km\,s^{-1}$ and resolution of $0.42\,\rm km\,s^{-1}$. 
We perform bandpass calibration via frequency switching, 
and all data were reduced using the Astronomical Image Processing
System\footnote{http://www.aips.nrao.edu} (AIPS).
The absorption spectra, $\tau(v)$, were extracted from the final cleaned data cubes
following calibration detailed in the 21\,SPONGE pilot paper \citep{Begum}.

In addition, for each sightline we obtain H\textsc{i} 
emission profiles which estimate the brightness 
temperature ($T_{\rm B}(v)$) in the absence of the radio 
continuum source.  Of our \NoOfSources{} sightlines, 11 
have emission profiles from the Millennium Arecibo 21-cm 
Absorption Line Survey \citep{HT031}, 10 have emission 
profiles from the Galactic Arecibo L-band Feed 
Array Survey in H\textsc{i} \citep[GALFA--H\textsc{i}; ][]{Stanimirovic06, GALFA}, 
and for 3 sightlines which were not included
in the Millennium survey or the GALFA-H\textsc{i} survey 
to date, we use emission spectra from the Leiden Argentine 
Bonn \citep[LAB; ][]{LAB} survey. Arecibo H\textsc{i} emission spectra 
have not been corrected for contamination
entering the telescope beam through distant sidelobes (so called stray radiation).

\begin{table*}
\centering
\caption{Source Information}
\begin{tabular}{l c c c c}
\hline
\hline
\small
 & $\sigma_\tau$\tablenotemark{a} & $W=1/\sigma_\tau$\tablenotemark{b}  & $l$ & $b$ \\ 
Name & ($\times10^{3}$) & ($\times10^2$) &(deg) & (deg) \\
\hline
4C32.44 & 1.5 & 2.7 & 67.240 & 81.049 \\ 
4C25.43 & 0.9 & 4.5 & 22.464 & 80.991 \\ 
3C286 & 0.7 & 5.8 & 56.527 & 80.676 \\ 
4C12.50 & 1.3 & 3.1 & 347.220 & 70.173 \\ 
3C273 & 0.6 & 6.7 & 289.945 & 64.358 \\ 
3C298 & 0.8 & 5.1 & 352.159 & 60.667 \\ 
3C225A & 1.5 & 2.7 & 219.866 & 44.025 \\ 
3C225B & 3 & 1.3 & 220.010 & 44.007 \\ 
3C345 & 1.3 & 3.1 & 63.455 & 40.948 \\ 
3C327.1 & 1.4 & 2.9 & 12.181 & 37.006 \\ 
3C147* & 0.4 & 10.1 & 161.686 & 10.298 \\ 
3C154* & 0.9 & 4.5 & 185.594 & 4.006 \\ 
3C410 & 1.9 & 2.1 & 69.210 & $-3.768$ \\ 
B2050+36* & 2.3 & 1.8 & 78.858 & $-5.124$ \\ 
P0531+19 & 0.5 & 8.1 & 186.760 & $-7.110$ \\ 
3C111* & 0.7 & 5.8 & 161.675 & $-8.821$ \\ 
3C133 & 1.8 & 2.2 & 177.725 & $-9.914$ \\ 
3C138 & 0.9 & 4.5 & 187.403 & $-11.347$ \\ 
3C123* & 0.7 & 5.8 & 170.581 & $-11.662$ \\ 
3C120 & 1.1 & 3.7 & 190.373 & $-27.397$ \\ 
3C48 & 0.7 & 5.8 & 133.961 & $-28.720$ \\ 
4C16.09 & 0.8 & 5.1 & 166.633 & $-33.598$ \\ 
3C454.3 & 0.9 & 4.5 & 86.108 & $-38.182$ \\ 
3C78 & 4.1 & 1.0 & 174.857 & $-44.514$ \\ 
\hline
\end{tabular}
\tablenotetext{*}{Excluded sources (see Section 2.2).}
\tablenotetext{a}{RMS noise in absorption profile calculated per $0.42\rm\,km\,s^{-1}$ channel.}
\tablenotetext{b}{Normalized weighting factor (see Section 3); $\Sigma_n W_n =1$.}
\end{table*}

\subsection{Derivation of $T_s$ for individual components}

To estimate $T_s$ for 
individual spectral components, we follow the method of \citet{HT031}.
We first fit each absorption profile with Gaussian functions 
assuming thermal and turbulent broadening. We apply the 
statistical $f$-test to determine the best-fit number of 
components.  
We then fit the $T_{\rm B}(v)$ profile simultaneously for
additional Gaussian components and $T_s$ for each absorption-detected 
component by solving the radiative transfer equations discussed at length in
\citet{HT031}.  These results will be presented along with 
additional survey information in Murray et al. 2014 (in prep).

We next construct ``residual'' spectra to search for weak absorption features 
below our observational sensitivity by removing fitted models 
from the absorption and emission profiles.
The best-fitting Gaussian components in absorption
($\tau_{i}(v)$ where $i$ denotes the $i^{\rm th}$ 
component) are subtracted from each original absorption profile $\tau(v)$ to produce a residual absorption spectrum,
$\tau_{\rm res} (v)= \tau(v) - \Sigma \tau_{i}(v)$.
The corresponding emission contribution from these components,
$\Sigma T_{s,i} (1-e^{-\tau_{i}(v)})$,
is removed from each emission profile, to produce a residual 
emission spectrum $T_{\rm B,res}(v) = T_{\rm B}(v) - \Sigma T_{s,i} (1-e^{-\tau_{i}(v)})$.
Each residual emission spectrum is dominated by signals from the WNM purely detected in emission.
Correspondingly, each residual absorption spectrum, $\tau_{res}(v)$,
contains noise, model imperfections and weak absorption
components below our detection threshold.

To minimize the effects of model imperfections in our results, we
exclude sources from our analysis which have strong
model-subtraction errors using the following technique.  We first
calculate the expected noise profile of the absorption spectrum,
 $\sigma_{\tau} (v)$, for each source.  Because the LAB survey
uses antennas of comparable size to those of the VLA, we follow methods
described by \citet{Roy13one} to estimate $\sigma_{\tau} (v)$ using
LAB H\textsc{i} emission.
We next  produce probability distribution functions
(PDFs) of $\tau_{res}(v)/\sigma_{\tau} (v)$ and  exclude from 
further analysis 5/24 sources whose PDFs deviate from a normal distribution at
$\geq97.5$\% confidence.
The excluded sources are significantly
contaminated by model-subtraction artifacts, and lie at
low Galactic latitude where lines of sight probe many
velocity-blended H\textsc{i} clouds, which complicates
Gaussian modeling. Table~1 lists the source name, $\sigma_{\tau}$
calculated per $0.42\rm\,km\,s^{-1}$ velocity channel in offline channels,
a weighting factor described in Section 3, and Galactic coordinates.

\section{Stacking analysis of HI absorption and emission spectra}
\label{s:stacking}

\begin{figure*}
    \centering
    \includegraphics[bb = 53 186 540 701, scale=0.7]{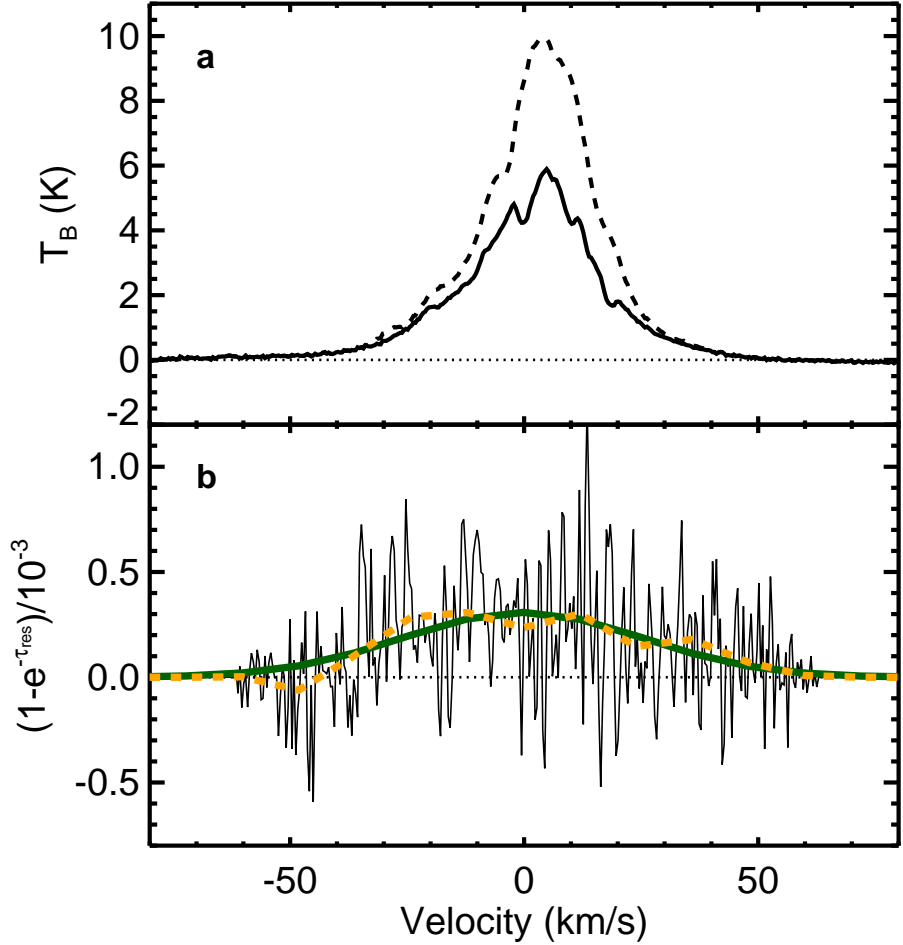}
    \caption{ (\textbf{a}) The
    	result of stacking the residual emission profiles $T_{\rm B, res}(v)$ (solid line) for the \NoOfKeeps{} selected sources, and the result of stacking the total emission profiles, $T_{\rm B}(v)$ (dashed
	line) for reference. (\textbf{b}) The result of stacking the residual VLA
         absorption profiles $\tau_{\rm res}(v)$ (thin solid black) from the selected \NoOfKeeps{} sources. A smoothed version 
         of the stack ($10\rm\,km\,s^{-1}$ boxcar kernel) is overlaid 
         (dashed orange), with a simple Gaussian fit to the profile 
         overlaid (thick solid green). Due to the fact that all shifted 
         profiles do not cover the same velocity range, we add zeroes to 
         the edges of each profile so that they cover the same range. 
         This results in fewer non-zero channels in the farthest 
         velocity bins, which causes the noise level to be lower there. }
    \label{f:stack}
\end{figure*}

We have performed a spectral ``stacking'' analysis on 
our 19 remaining residual spectra to search for 
extremely weak absorption signals from the 
diffuse WNM. 
We first apply a velocity shift to both the residual emission and absorption profiles
to remove the effect of Galactic rotation and align any remaining 
signals at $0\,\rm km\,s^{-1}$.  The velocity shifts are computed 
using the first velocity moment of the residual emission spectrum for each source, so that 
$\Delta v=-\int T_{\rm B,res}\,v\,dv/\int T_{\rm B,res} dv$.
To maximize the signal to noise ratio of the stacked absorption spectrum, 
the weight, $W$, for each profile is given by
$W = \tau_{\rm res}/\sigma_{\tau}^2$
(Treister et al. 2011).
However, as discussed in the previous section,
$\sigma_{\tau} \propto T_{\rm B}$.  For constant $T_s$, we have
$\tau_{\rm res} \propto T_{\rm B}$, and so the weight simplifies to:
$W=1/\sigma_{\tau}$. We measure $\sigma_{\tau}$
in the offline channels of each profile, and list these values with
the weights $W$, normalized by the sum over all $n=19$ profiles,
$\Sigma_{n} W_{n}$, in Table 1. The same weighting 
values were applied to the residual emission spectra.

The weighted profiles are averaged to produce the final stacked emission 
and stacked absorption spectra shown in Figure \ref{f:stack}.  The RMS 
noise in the stacked absorption spectrum is  \RmsStack{}, calculated over a fixed range of channels 
($20$ to $30$ and -$30$ to -$20\rm\,km\,s^{-1}$). This is \TimesBetter{} times more 
sensitive than the median RMS noise calculated in the same 
velocity channels in the individual residual absorption 
profiles ($\sigma_{\tau}=1.15\times10^{-3}$).

The enhanced sensitivity enabled by stacking allows 
us to detect a weak and broad absorption component which has 
a velocity width and centroid consistent with the 
stacked {\em emission} signal (Figure~\ref{f:stack}). This 
gives confidence that the stacked profiles trace physically 
related quantities. This broad absorption component is 
the weighted-mean, $0\,\rm km\,s^{-1}$-centered absorption signal 
over all sightlines, and has an equivalent width ($\rm EW$) 
within the velocity range common to all shifted profiles 
($-$46 to $30\rm\,km\,s^{-1}$) of 
${\rm EW}=\int_{-46}^{30} (1-e^{-\tau_{\rm res}}) d v$ = \EW{}. 

\begin{figure*}
    \centering
    \includegraphics[bb = 18 425 568 739, scale=0.7]{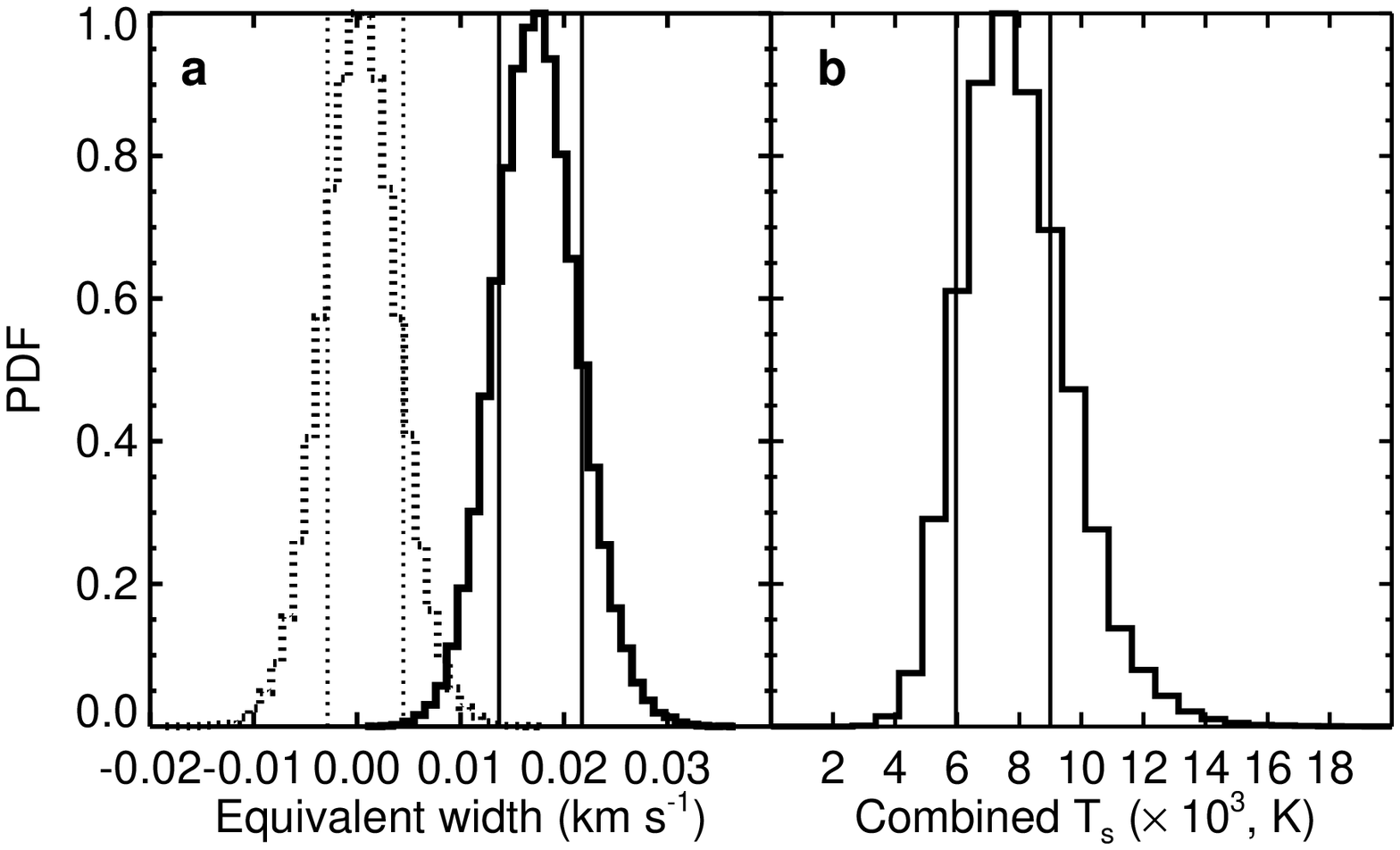}
    \caption{(\textbf{a}) PDF of stacked absorption 
         $\rm EW$ following \NoIterations{} iterations using a 
         bootstrapping algorithm to select \NoOfKeeps{} sources 
         from the  total list of \NoOfKeeps{} with replacement, 
         yielding \MonteCarloEW{} ($68$\% confidence). To test 
         that the signal is real, we invert (or multiply by $-1$) 
         a random selection of half of the absorption residual 
         profiles in \NoIterations{} additional trials, yielding 
         a distribution (dotted histogram) consistent with zero 
         signal, as expected. Vertical lines denote $1\sigma$ 
         uncertainty limits. (\textbf{b}) Distribution of spin 
         temperatures from combining the PDFs of three different 
         estimations (see text) computed in all \NoIterations{} 
         bootstrapping trials.  The derived spin temperature is 
         \StackedSpinTemp{} ($68$\% confidence), with a $99\%$ 
         confidence lower limit of \LowLimit{}.
         }
    \label{f:histo}
\end{figure*}

We next conduct a bootstrapping \citep[see, e.g., ][]{Wall03}
Monte Carlo simulation to estimate the integrated strength of 
the stacked absorption signal (Figure~\ref{f:stack}b) and test 
for contamination by outlier spectra.  We run the stacking 
analysis on a new sample of \NoOfKeeps{} sightlines randomly 
chosen from our original \NoOfKeeps{} sightlines with 
replacement.  We repeat this trial \NoIterations{} times, 
each time recomputing the $\rm EW$.  The resultant normalized 
PDF of $\rm EW$s (Figure~\ref{f:histo}a, solid histogram) 
is nearly Gaussian, showing that the stacked $\rm EW$ 
signal is consistent with being drawn from a parent 
population of spectra with comparable means, and not 
due to a few outlier spectra. The peak of the $\rm EW$ 
distribution and the numerically-integrated $68\%$ 
confidence limits are \MonteCarloEW{}, giving 1 in 
$2\times10^6$ chance of being spurious (\StackedSNR{}$\sigma$).
Figure~\ref{f:histo}a also displays the result of repeating 
the bootstrapping simulation while inverting (or multiplying 
by $-1$) a random selection of half of the profiles (dotted 
histogram). The result is fully consistent with zero signal, 
verifying that our stacking method does not produce 
spurious detections.

Using the Monte Carlo bootstrapping method, we also 
constrain the FWHM and peak optical depth of the stacked 
absorption feature by Gaussian fit as \MCfwhm{} and \MCpeak{} (at the central
velocity of the feature, $v_0$), 
respectively. Both quantities have well-defined, 
single-peaked distributions, suggesting that the detection 
is tracing a single gas component rather than a blend of 
many narrow components.  We therefore proceed to estimate 
the typical spin temperature $T_s$ of the gas detected 
in the stacked absorption spectrum.

We use three methods: (1) 
$T_s = \int_{-46}^{30} T_{\rm B, res}\, dv /
\int_{-46}^{30} (1-e^{-\tau_{\rm res}})\,dv$, (2) 
$T_s = T_{\rm B, res}(v_o)/(1-e^{-\tau_{\rm res}})(v_o)$, 
where $v_o$ is the velocity at the peak of the smoothed 
stacked absorption signal, and (3) by the 
methods of \citet{HT031}, fitting a single Gaussian 
component to the absorption stack and solving for $T_s$ 
by fitting this component to the emission stack without additional
components. For each of the \NoIterations{} bootstrapping 
trials, we compute all three $T_s$ estimates and combine 
their distributions to minimize systematic errors 
associated with any individual approach 
(Figure~\ref{f:histo}b).

We estimate a spin temperature of the detected  
absorption feature of \StackedSpinTemp{} ($68$\% confidence), 
with a lower limit at $99$\% confidence of \LowLimit{}.
We note that by shifting the residual profiles by the 
second velocity moment or by the location of maximum 
$T_{\rm B,res}$, we find consistent temperature estimates 
with similar significance. 
The possible contamination of stray radiation to 
the Arecibo H\textsc{i} emission spectra
has no effect on the presence of the absorption
stack, although it will tend to increase the EW of the
emission stack, and therefore increase the estimated
$T_s$  of the stacked feature. For the GALFA-H\textsc{i} 
survey, \citet{GALFA} estimated that the level of stray radiation contamination
is $\leq 200$--$500\rm\,mK$, leading to an overestimate in
our computed $T_s$ of at most 13\%. 
Therefore, the presence of stray radiation in our Arecibo HI emission spectra would
not change our results

\section{Discussion}
\label{s:discussion}

\begin{figure*}
    \centering
    \includegraphics[scale=0.5,angle=90]{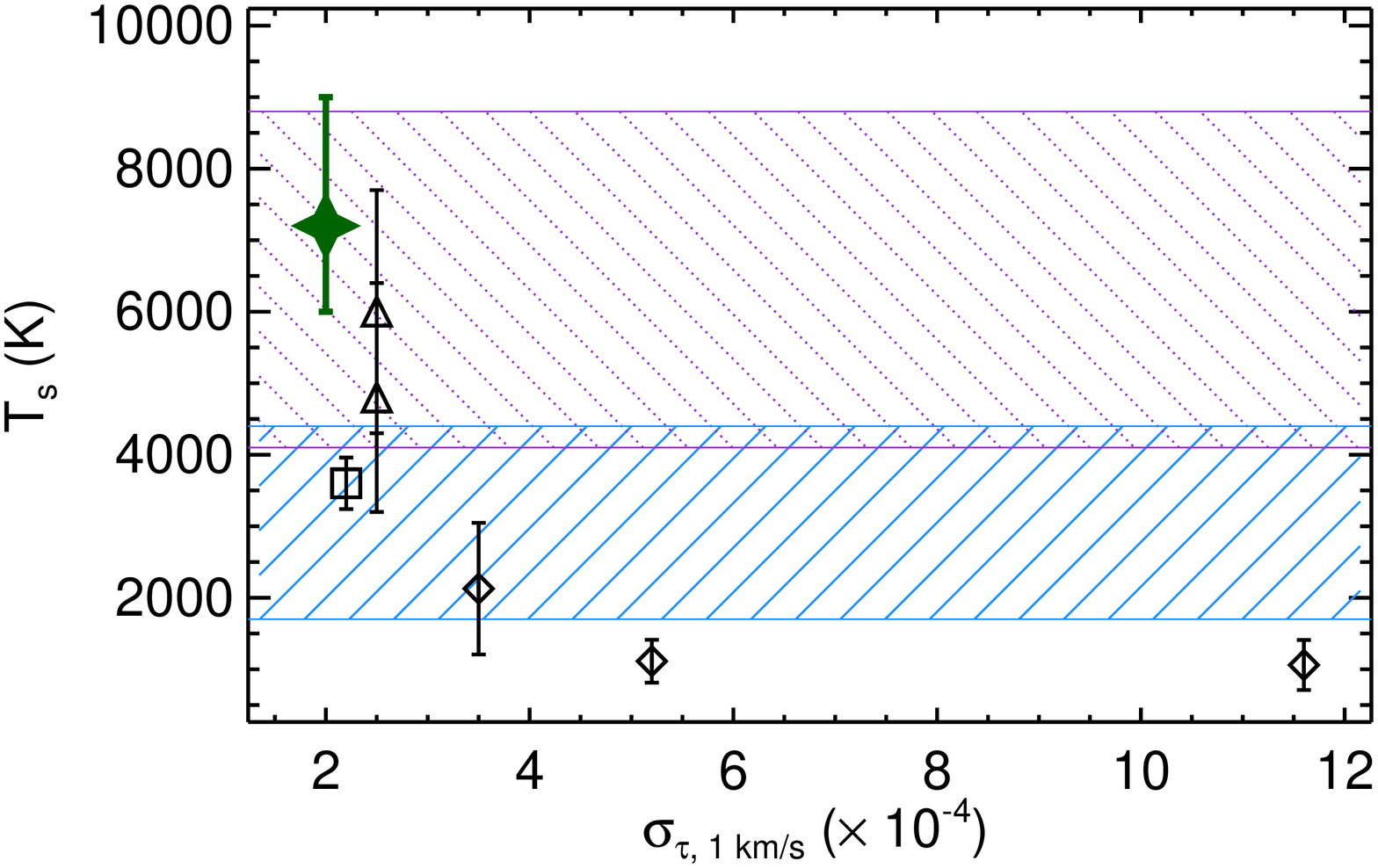}
    \caption{Comparison of previous $T_s$ 
         measurements for the WNM. Hollow symbols denote direct 
         observational measurements of individual sightline 
         components. \emph{Triangles}: Absorption towards Cygnus A 
         \citep{CDG98}, \emph{Square}: Absorption towards 
         3C147 \citep{DCG02} \emph{Diamonds}: 
         21\,SPONGE absorption towards PKS0531+19, 3C298 and 3C133.  All 
         points are plotted versus the RMS noise in off-line channels 
         in the absorption profile, per $1\rm km\,s^{-1}$ channel 
         ($\sigma_{\tau,1\rm km/s}$). The purple dotted hatched region 
         denotes the \emph{kinetic} temperature range from 
         \citet{Wolfire03} ($T_k\sim4100$--$8800\rm\,K$), and the 
         blue hatched region denotes the \emph{spin} temperature 
         range from \citet{Liszt01} for all 
         possbile ISM pressures ($T_s\sim1800$--$4400\rm\,K$). The 
         green filled star is the derived spin temperature from 
         Figure~\ref{f:histo}b, \StackedSpinTemp{} ($68$\% confidence),  with a 
         $99\%$ confidence lower limit of \LowLimit{}.}
    \label{f:temp}
\end{figure*}

In Figure~\ref{f:temp}, we plot all previous WNM $T_s$ detections 
(open symbols) against the RMS noise in off-line channels of the 
individual absorption spectra in which they were detected, with 
$1\sigma$ error bars derived from line fitting.  We exclude 
literature measurements which are estimated as upper limits 
from $T_k$ or which assume single temperatures for  
multi-component sightlines.  Our result from the stacking 
analysis (green filled star) is in agreement with 
measurements by \citet{CDG98} (open triangles) obtained in the 
direction of $\sim 400\rm\,Jy$-bright Cygnus A.  We emphasize 
that our result samples a widespread population detected 
from \NoOfKeeps{} sightlines,  while \citet{CDG98} examine 
only a single direction. Our $T_s$ estimate is significantly 
higher than all other direct measurements shown here. The 
trend of increasing $T_s$ with increasing observational 
sensitivity confirms the expectation that previous experiments 
with lower sensitivity ($\sigma_{\tau}\geq5\times10^{-4}$)
were unable to detect WNM with $T_s>1000\rm\,K$ due to its 
low optical depth.

The range of predicted \emph{kinetic} temperatures from the 
most detailed ISM heating and cooling considerations, shown by
the dotted purple hatched region in Figure~\ref{f:temp}, 
is $T_k\sim 4100$--$8800\rm\,K$ \citep{Wolfire03}. 
We use models considering 
only collisional excitation from \citet{Liszt01} (their Figure 2) to show that
under all plausible ISM pressures, 
the range of spin temperatures implied by this $T_k$
range from \citet{Wolfire03}, is $T_s\sim 1800$--$4400\rm\,K$ (solid blue hatched 
region in Figure~\ref{f:temp}).
Our measurement indicates that the mean spin 
temperature of the WNM is higher than expected theoretically 
for collisionally excited H\textsc{i} at 98\% confidence.  

Resonant scattering of Ly-$\alpha$ photons can contribute enough
to 21-cm excitation to allow  $T_s=T_k$, as long as a sufficient
fraction of the Ly-$\alpha$ radiation permeates the WNM 
\citep{Liszt01,Pritchard12}.  However, the degree to which 
this can occur in a multi-phase ISM depends on several 
observationally unconstrained quantities: the local and 
external Ly-$\alpha$ field, interstellar pressure, interstellar 
turbulence, ionization fraction, and the topology of the ISM. 
Using models from \citet{Liszt01}, which assume a column 
density of hydrogen nuclei equal to $10^{19}\rm\,cm^{-2}$ and 
the temperature of the Ly-$\alpha$ radiation field being 
equal to the kinetic temperature, our $T_s$ measurement 
constrains the fraction of Galactic flux from early-type 
stars which permeates H\textsc{i} clouds to $> 1\times 10^{-4}$.

Our work provides the first observational evidence for the Ly-$\alpha$ 
mechanism acting throughout the bulk of the Galactic WNM. 
By increasing observational sensitivity to $T_s$ in 
comparison with expectations for $T_k$, stacking can be 
used to constrain the importance of non-collisional 
excitation on H\textsc{i}, as well as the origin and intensity 
of the Ly-$\alpha$ radiation field.  In the future, 
21\,SPONGE will obtain more absorption sightlines to 
increase our sensitivity to weak underlying signatures of 
the WNM, thereby further refining these temperature 
constraints, and sampling different Galactic environments.

\section{Conclusions}
\label{s:conclusions}

We have presented the discovery of a weak 
(\MCpeak{}), broad (\MCfwhm{}) WNM absorption 
feature in the stacked absorption 
spectrum of \NoOfKeeps{} independent Galactic 
sight-lines from the 21-SPONGE survey.  Using 
Monte Carlo simulations, we have estimated the 
feature's spin temperature to be \StackedSpinTemp{}, 
which is significantly (98\% confidence) higher 
than theoretical predictions based on collisional 
excitation alone, likely due to the
thermalization of H\textsc{i} by resonant 
Ly-$\alpha$ scattering.
This work provides the first observational evidence that the 
Ly-$\alpha$ excitation mechanism is acting 
throughout the bulk of the Galactic WNM, and 
demonstrates that the Ly-$\alpha$ radiation 
field, a quantity that is difficult to measure yet
vitally important for interpreting H\textsc{i} 
signals from early epochs of cosmic structure 
formation, can be probed using measurements of the WNM.

\acknowledgements
We thank an anonymous referee for helpful comments and suggestions.
This work was supported by the NSF Early Career 
Development (CAREER) Award AST-1056780. C.~M. acknowledges 
support by the National Science Foundation Graduate Research 
Fellowship and the Wisconsin Space Grant Institution. 
S.~S. thanks the Research Corporation for Science 
Advancement for their support.  We thank A. Begum for 
help with initial survey observations, D. Able for developing
initial data reduction strategy, and H. Liszt, J. S. Gallagher, N. Roy
and T. Wong for helpful discussions. The National Radio 
Astronomy Observatory is a facility of the National Science 
Foundation operated under cooperative agreement by Associated 
Universities, Inc. The Arecibo Observatory is operated by 
SRI International under a cooperative agreement with the 
National Science Foundation (AST-1100968), and in alliance 
with Ana G. M\'endez-Universidad Metropolitana, and the 
Universities Space Research Association.

\bibliographystyle{apj}

\label{lastpage}
\end{document}